\documentclass{INTERSPEECH2023}

\interspeechcameraready

\title{Towards Spontaneous Style Modeling with Semi-supervised Pre-training for Conversational Text-to-Speech Synthesis}
\name{Weiqin Li$^{1,*}$\thanks{* Work conducted when the first author was intern at XVerse Inc.}, Shun Lei$^{1}$, Qiaochu Huang$^{1}$, Yixuan Zhou$^{1}$, \\Zhiyong Wu$^{1,3,\dagger}$\thanks{$\dagger$ Corresponding author.}, Shiyin Kang$^{2}$, Helen Meng$^{3}$
}
\address{
  $^1$Shenzhen International Graduate School, Tsinghua University, Shenzhen, China\\
  $^2$XVerse Inc., Shenzhen, China\\
  $^3$The Chinese University of Hong Kong, Hong Kong SAR, China}
\email{\{lwq22, leis21\}$@$mails.tsinghua.edu.cn, zywu$@$sz.tsinghua.edu.cn}
\begin{document}

\maketitle
\begin{abstract}
The spontaneous behavior that often occurs in conversations makes speech more human-like compared to reading-style.
However, synthesizing spontaneous-style speech is challenging due to the lack of high-quality spontaneous datasets and the high cost of labeling spontaneous behavior.
In this paper, we propose a semi-supervised pre-training method 
to increase the amount of spontaneous-style speech and spontaneous behavioral labels.
In the process of semi-supervised learning, both text and speech information are considered for detecting spontaneous behaviors labels in speech.
Moreover, a linguistic-aware encoder is used to model the relationship between each sentence in the conversation.
Experimental results indicate that our proposed method achieves superior expressive speech synthesis performance with the ability to model spontaneous behavior in spontaneous-style speech and predict reasonable spontaneous behavior from text.
\end{abstract}
\section{Introduction}
Text-to-speech (TTS) aims to synthesize intelligible and natural speech from text~\cite{survey,tacotron,fastspeech}.
With the development of deep learning, TTS models are able to synthesize reading-style speech exceptionally well~\cite{tacotron2, deepvoice3, fastspeech2}.
However, existing TTS systems are unable to provide sufficient performance and an immersive experience for spontaneous-style speech which typically occurs in conversations. 
Conversations often contain subtle spontaneous behaviors such as filled pause, prolongation, repair, repetition, laughing, coughing, etc., that make speech sound more genuine~\cite{filledpause,spontaneousClassify,repairAndRepetition,prolongation}.
Hence, the modeling of spontaneous behaviors play a crucial role in spontaneous speech synthesis.

Some studies have applied existing TTS systems to synthesize spontaneous-style speech.
In~\cite{converTTS}, the syntactic structure and chat history have proven effective for the modeling of spontaneous style in conversation.
Some early studies found that using explicit labels to represent two typical spontaneous behaviors, filled pause and prolongation, could help the TTS model generate spontaneous-style speech and allow control of spontaneous behaviors by adding labels~\cite{fpInTTS,controllabelSpon}.
Besides, modeling some spontaneous behaviors was shown to increase the naturalness of spontaneous-style speech, such as breathing~\cite{breathingSpontaneous}, laughter~\cite{latenStyleSpontaneous}, interjections~\cite{interjections}, and creaky phonation~\cite{creaky}.

The lack of high-quality spontaneous dataset and the high cost of manually labeling spontaneous behaviors constraints to the development of spontaneous-style conversation TTS.
A strategy for constructing a dataset of spontaneous-style conversations is described in~\cite{converTTS}, but the cost of recording and labeling is expensive.~\cite{Adaspeech3} proposed a spontaneous dataset mining method to collect spontaneous datasets from podcasts.
This method reduces the cost of collecting speech, but it assumes in advance that spontaneous behaviors appear only on specific words, which contradicts the diversity of spontaneous behaviors and the randomness of their appearance.
In tasks with limited labeled data, semi-supervised learning methods for extracting pseudo labels from large-scale datasets are shown to be useful for augmenting the data~\cite{semi-supervised}.

Moreover, it is difficult to model the spontaneous style in conversation simply based on the current utterance.
Previous work used historical conversation information to enhance spontaneous-style speech synthesis~\cite{converTTS}.
However, the linguistic information in conversation are not well used, which includes the relationship between individual sentences in the conversation.
In~\cite{paratts}, linguistic information from the paragraph was utilized to increase the expressiveness of the synthesized speech.
In addition, the prediction of spontaneous behavior labels in previous works~\cite{controllabelSpon,Adaspeech3} was implemented in the text frontend, resulting in unnatural generated speech when unreasonable labels were present.

In this paper, we use a linguistic-aware spontaneous conversational speech synthesis model based on Fastspeech 2~\cite{fastspeech2} to generate spontaneous-style speech in conversation.
To address the lack of high-quality labeled spontaneous datasets, we propose a semi-supervised pre-training method conducted on a large-scale low-quality spontaneous dataset to increase the amount of spontaneous-style speech and spontaneous behavioral labels.
In the process of semi-supervised learning, text and speech information are considered to detect spontaneous behaviors in speech.
Inspired by~\cite{paratts}, we introduce a linguistic-aware encoder to model the linguistic information in the conversation.
Experiments results show that our proposed method enables the model to synthesize more natural spontaneous-style speech. The use of a semi-supervised pre-training method substantially improves the model's ability to predict reasonable labels, while the addition of a linguistic-aware encoder improves the model's capacity to simulate spontaneous behavior in speech.

\section{Methodology}

\begin{figure}[!tb]
	\centering
	\includegraphics[width=0.8\linewidth, height=1.1\linewidth]{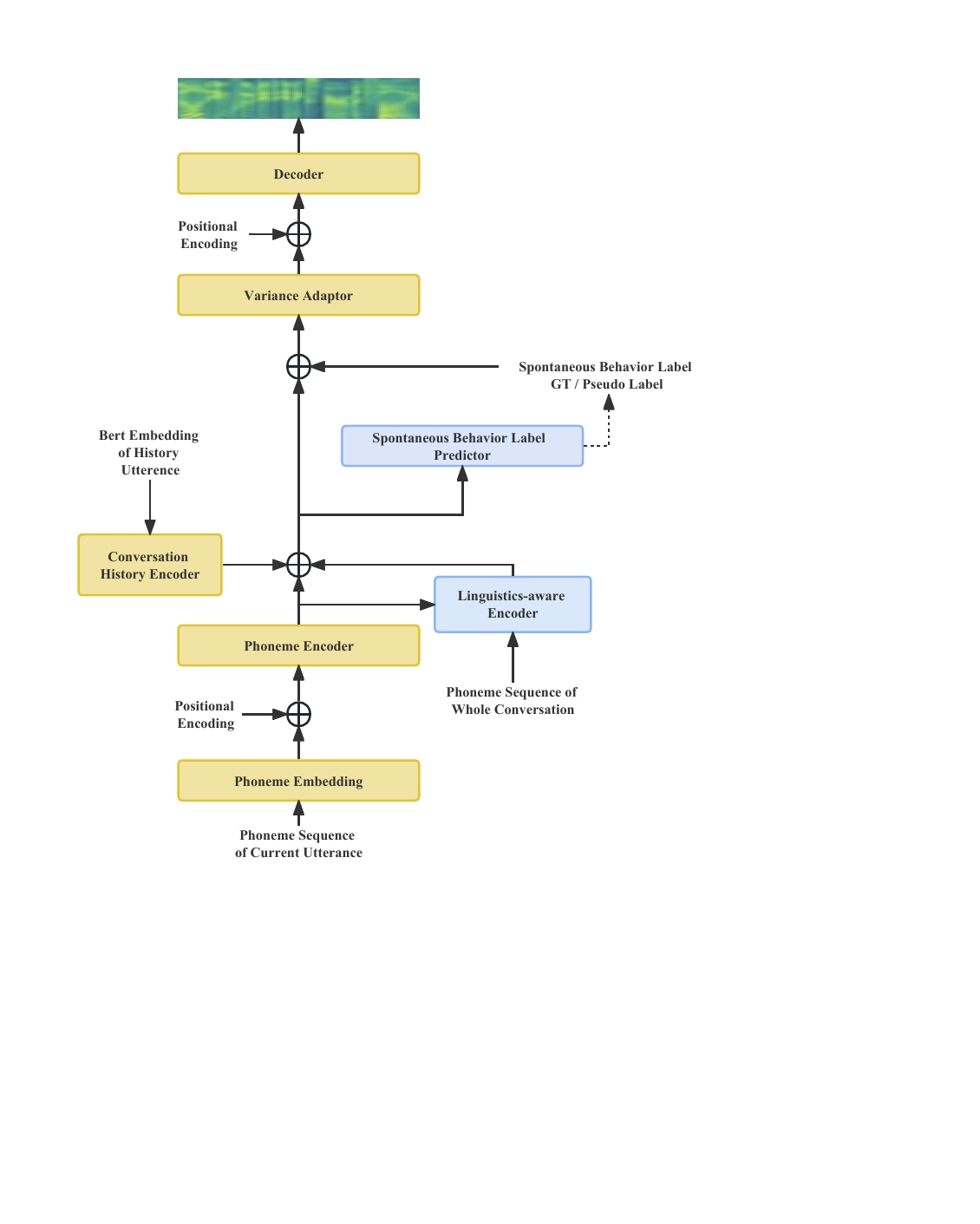}
	\caption{The architecture of our proposed model. For inference phase, the output of the spontaneous behavior label predictor is used. Pseudo label is only used in pre-training phase.}
	\label{fig:struct_new}
        \vspace{-1.0em}
\end{figure}

The architecture of our proposed model is illustrated in Fig.\ref{fig:struct_new}.
As the backbone of the TTS model, Fastspeeh 2~\cite{fastspeech2} and the conversation history encoder are combined to generate mel-spectrograms from a given phoneme sequence.
For modeling and predicting spontaneous behavior, we use a label detector (Fig.\ref{fig:PseudoLabelPredictor}) to extract pseudo labels and introduce a spontaneous behavior label predictor to model and predict the spontaneous behavior.
Besides, a linguistic-aware encoder is proposed to model linguistic information in conversations.
We describe the important parts of the model and the method of pre-training and fine-tuning in the following subsections.


\vspace{-0.5em}
\subsection{Backbone Framework}
\vspace{-0.35em}
We follow the previous work on spontaneous conversational speech synthesis~\cite{controllabelSpon} and make some modifications as our backbone, as shown in Figure.\ref{fig:struct_new}.
Firstly, Fastspeech 2~\cite{fastspeech2} replaces Tacotron 2~\cite{tacotron2} as an acoustic model to improve the speed and robustness of speech synthesis.
Secondly, the process of predicting spontaneous labels is implemented within the TTS model as opposed to independently in the text frontend, reducing the cumulative effect of prediction errors on the generated speech.
Thirdly, we not only consider the bert embedding of current utterance, but also use the bert embedding of historical utterances to obtain richer semantic information in historical conversation.
The conversation history encoder in~\cite{converTTS} is used to model the semantic information.
Specifically, the utterance-level bert embedding of the current utterance and the previous 5 utterances is extracted using a pre-trained bert model, and then fed into the encoder consisting of the Gated Recurrent Unit (GRU)~\cite{GRU} and linear layers to generate the output, which is added to the output of the phoneme encoder.

\vspace{-0.8em}
\subsection{Spontaneous Behavior Modeling and Predicting}
For modeling spontaneous behavior, our model focuses on two frequent spontaneous behaviors: filled pause and prolongation. Specifically, a filled pause is a semantically empty element of speech that delays the transfer of the speaker's message and is usually expressed in the form
of "em", "uh", etc~\cite{filledpause}. 
Prolongations are mainly used to indicate hesitation and to emphasize the discourse focus~\cite{prolongation}.
Since they both come after a character, we express them with an explicit label at the character level.
\vspace{-0.3em}
\subsubsection{Multi-Model Pseudo Label Detector}
\begin{figure}[!tb]
	\centering
	\includegraphics[width=0.95\linewidth, height=0.25\linewidth]{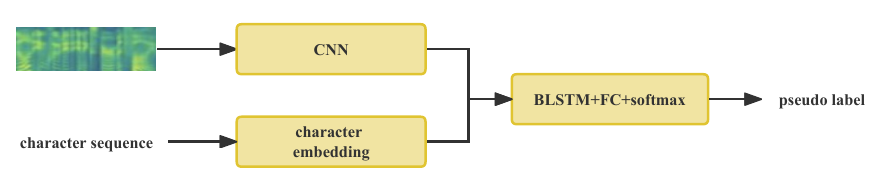}
	\caption{The structure of the multi-model pseudo label detector.}
	\label{fig:PseudoLabelPredictor}
        \vspace{-1.0em}
\end{figure}

\vspace{-0.3em}
To extract pseudo labels from a low-quality, unlabeled spontaneous dataset~\cite{RAMC}, we construct a multi-modal pseudo label detector with a structure similar to~\cite{labeldetector}.
It is difficult to adequately detect spontaneous behavior from speech based on acoustic features alone.
Since the appearance of spontaneous behavior in speech is impacted by the semantics of the text and is strongly related to the occurrence and position of some specific words,
text can be introduced as auxiliary information to help detect spontaneous behavior in speech.
Thus, this detector employs textual and acoustic features to detect spontaneous behavior labels.
The structure is shown in Fig.\ref{fig:PseudoLabelPredictor}.

The mel-spectrum is downsampled by the Convolutional Neural Network (CNN) in time and channel dimensions, and then the results are transformed from frame level to character level by summation. The obtained character-level acoustic features are concatenated with the character embedding of the text and fed to a detector comprised of a Bi-directional Long Short-Term Memory (BLSTM) and fully connected layer. Finally, the softmax result is achieved, and then the character-level predicted result is obtained according to the set threshold value. Since inappropriate spontaneous behavior diminishes the naturalness of the generated speech, we increase the threshold to improve the precision of the detection model.

\vspace{-0.6em}
\subsubsection{Spontaneous Behavior Label Predictor}
\vspace{-0.35em}
Typically, there are no explicit spontaneous labels in the process of speech synthesis. To automatically predict spontaneous behavior from text representations and thus make the generated speech more human-like, we construct a spontaneous label predictor after the phoneme encoder.
The structure of the label predictor is the same as the variance predictor in Fastspeech 2~\cite{fastspeech2}. 
This predictor takes the phoneme hidden sequence as input and predicts the spontaneous labels for each phoneme.
We combine the two spontaneous behavior labels: 0 for none, 1 for filled pause, 2 for prolongation, and 3 for both happening.
The character-level spontaneous label sequence is expanded to the phoneme-level by simple repetition, and only the last phoneme of a character has a positive label.

During training, ground truth spontaneous labels are utilized to train the predictor and label embedding, and the loss function is MSE loss. In inference, it is possible to supply explicit labels or not, and when labels are not provided, the output of the predictor is used.
Finally, the label embedding output is added to the text hidden representation and supplied to the variance adaptor for predicting duration, pitch, and energy.

\vspace{-0.6em}
\subsection{Linguistics-aware Encoder}
Conversations contain a wealth of linguistic information, including the relationship between individual sentences. The modeling of linguistic information can enhance the expressiveness of synthetic spontaneous behavior.

\begin{figure}[!tb]
	\centering
	\includegraphics[width=0.8\linewidth, height=0.71\linewidth]{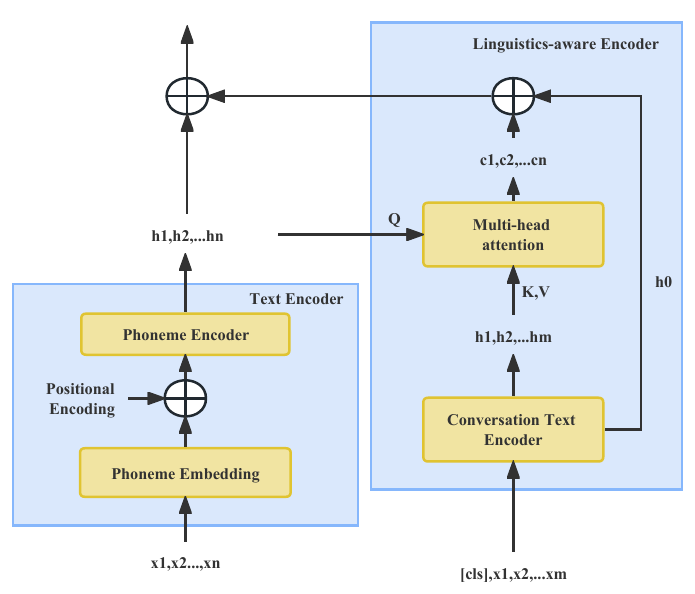}
	\caption{The structure of the linguistics-aware encoder.}
	\label{fig:lingusticEncoder}
        \vspace{-0.3em}
\end{figure}

To better model the linguistic information in the conversations, we introduced a linguistic-aware encoder consisting of a conversation encoder with a multi-head attention mechanism~\cite{attention}, as shown in Fig.\ref{fig:lingusticEncoder}.
The conversation text encoder has the same structure as the text encoder in Fastspeech 2 \cite{fastspeech2}.

In details, we are given a conversational phoneme sequence of length m, \textbf{${x_c}$}$=(x_1,x_2,\ldots,x_m)$, and an utterance phoneme sequence of length n, \textbf{${x_u}$}$=(x_1,x_2,\ldots,x_n)$.
They are passed through the text encoder to obtain the hidden representations \textbf{${h_c}$}$=(h_1,h_2,\ldots,h_m)$ and \textbf{${h_u}$}$=(h_1,h_2,\ldots,h_n)$.
As for attention mechanisms, we use multi-head attention mechanisms~\cite{attention} to obtain linguistic connections between different independent sentences, which set $h_u$ as the query and $h_c$ as the key and value. The output of attention provides linguistic information about the current utterance and other utterances in the conversation.

To generate a linguistic representation of the entire conversation, we add a CLS tag to the conversational phoneme sequence in the input and represent the conversation linguistic representation with the corresponding high-level representation $h_0$. The outputs of the multi-head attention and the $h_0$ are added and used as the output of the linguistic-aware encoder. Finally, we add the output and the encoder output of the TTS backbone model to make the model aware of the linguistic knowledge related to conversations.

\vspace{-0.3em}
\subsection{Pre-training and Fine-tuning}
Although the prediction of spontaneous behavior can be achieved by the spontaneous behavior label predictor as described in section 2.2.2, it is challenging to predict reasonable spontaneous behavior because there are few high-quality spontaneous speech datasets with labels.
To augment the spontaneous dataset and spontaneous behavior labels used for training, we propose a semi-supervised pre-training method conducted on a large-scale low-quality spontaneous dataset.
Specifically, the training process of the model consists of the following four steps:
1) Training detector.
We train independent detection models for the two spontaneous behaviors using a high-quality, labeled spontaneous dataset.
2) Pseudo label extraction. 
Based on the trained behavior detector, we extract character-level pseudo labels of spontaneous behaviors from the RAMC dataset~\cite{RAMC}.
Pseudo labels generated by this semi-supervised approach can increase the amount of labels used for model training.
3) Pre-training.
The TTS model is then pre-trained using low-quality speech and pseudo labels to learn the mapping between text and spontaneous behavior labels.
4) Fine-tuning.
Finally, a high-quality spontaneous dataset is used to fine-tune the model.
In order to preserve the ability of the model to predict spontaneous labels based on textual information without learning the noise of low-quality datasets, we reset the decoder's parameters and increased the learning rate of the decoder by a factor of 10 during the fine-tuning phase.
\section{Experiments}

\subsection{Basic Setup}
We conduct our experiments on two Mandarin conversation corpus.
For pre-taining, we use an open source corpus called RAMC~\cite{RAMC}, which contains roughly 180 hours of spontaneous conversational speech with 351 groups of multi-round Mandarin conversations by 663 Mandarin speakers from different regions.
For training detector and fine-tuning, we use an internal high-quality corpus contains 6 hours of conversational speech consisting of 486 conversations recorded by two female speakers, where there are about 4,294 spontaneous labels for filled pause and prolongation. We randomly selected 36 conversations with a total of 359 sentences as the test set for both detection model and TTS model.

For feature extraction, we first resample all speech to 22.05 kHz, and then 80-dimensional mel-spectrograms are extracted from speech.
The filter length is set to 1,024 and the hop length is set to 256.
The phoneme duration is extracted by Montreal Forced Aligner \cite{mfa} tool.
An open-source pre-trained BERT-base model\footnote{\href{https://github.com/UKPLab/sentence-transformers}{https://github.com/UKPLab/sentence-transformers}\label{sentenBert}}~\cite{reimers2019sentence} is used in our experiments to extract 512-dim utterance-level BERT embedding.

In the training detector phase, we train 50 epochs and the loss function uses CrossEntropyLoss.
The threshold values for the filled pause detector and prolongation detector are set to 0.85 and 0.95. 
For the TTS model, we take 300k iterations for pre-training and 150k iterations for fine-tuning.
The Adam optimizer is adopted with $\beta_1=0.9$, $\beta_2=0.98$, $\epsilon=10^{-9}$ and the warm-up strategy is employed before 4000 iterations.
In addition, we employ a well-trained HiFi-GAN\cite{kong2020hifi} as the vocoder.

\subsection{Compared Methods} 
Two FastSpeech 2 based models are implemented for comparison, and the details are described as follows:

\textbf{FastSpeech 2}
An open-source implementation\footnote{Implemented based on:
\href{https://github.com/ming024/FastSpeech2}{https://github.com/ming024/FastSpeech2}\label{fn_fs2}} of vanilla FastSpeech 2 \cite{fastspeech2} which the input is phoneme sequence without explicit spontaneous labels.

\textbf{UCS*}
Unified controllable spontaneous conversational speech synthesis (UCS) model. 
We made some modifications to the UCS and used it as our backbone framework, as described in section 2.1.

\textbf{Proposed} 
The model we propose in this paper, which adopt the semi-supervised pre-training method and the linguistics-aware encoder.

\subsection{Objective Evaluation}
We use objective evaluation to measure the performance of label detection models. The precision, recall and F1-score are used as the metrics for objective evaluation. 
As shown in Table~\ref{tab:objective}, we compared the performance of label detectors with different input types.
Combining text and speech as inputs to the detector significantly improves the model's recall in comparison to speech-only input, demonstrating that
text information can help detect some reasonable spontaneous behavior labels that are difficult to get from speech information.
In addition, the addition of textual information improves the model's precision, hence enhancing the reasonableness of the spontaneous labels predicted by the model.
When using text and speech for input, the detectors for both spontaneous behaviors have a precision above 0.8 and a recall above 0.6, indicating that the models can detect enough amount of reasonable spontaneous behavior labels to achieve the purpose of expanding the number of labels in low-quality spontaneous datasets. 

\begin{table}[th]\footnotesize
\renewcommand{\arraystretch}{1}
  \caption{Objective evaluation of detection models with different spontaneous behaviors and different input types.}
  \label{tab:objective}
  \centering
  \begin{tabular}{lcccc} 
    \toprule
    \textbf{Input type} & \textbf{Behavior type} & \textbf{Precision} & \textbf{Recall} & \textbf{F1-score}\\
    \midrule
    speech  &filled pause  & $0.815$ & $0.524$ & $0.638$~~  \\
    speech  &prolongation & $0.8$ & $0.557$ & $0.657$~~ \\
    text+speech  &filled pause  & $0.866$ & $0.619$ & $0.722$~~  \\
    text+speech  &prolongation & $0.844$ & $0.710$ & $0.771$~~ \\
    \bottomrule
  \end{tabular}
\end{table}

Following~\cite{controllabelSpon}, the goal of predicting labels is to find some reasonable spontaneous positions to make the generated speech more natural. Hence, the performance of spontaneous label prediction is directly evaluated by subjective evaluation.

\subsection{Subjective Evaluation}
To evaluate the ability of the model to generate natural spontaneous-style speech, we conduct two mean opinion score (MOS) test.
The first MOS tests provides explicit spontaneous behavior labels during the inference phase, focusing on the naturalness of spontaneous behavior in generated speech.
The second MOS test does not provide labels in the inference phase but uses labels predicted by the model from the text. It mainly evaluates the reasonableness of the spontaneous behavior predicted by the model.
As neither Fastspeech 2 nor Ground Truth consider spontaneous behavior labels, the inference process for both MOS tests is identical.
We randomly selected 20 sentences from the test set, 30 native Mandarin speakers are recruited to evaluate the speeches on a scale from 1 to 5.
As shown in Table \ref{tab:mos}, the outcomes demonstrate the efficacy of our proposed methods in comparison to the baselines.
Our proposed approach achieves the best MOS of 3.561 in the test with labels, exceeding FastSpeech 2 by $0.716$ and UCS* by $0.2$.
The proposed method also obtains the best MOS of 3.701 in the test without labels, exceeding FastSpeech 2 by $0.796$ and UCS* by $0.28$.

\begin{table}[th]\footnotesize
\renewcommand{\arraystretch}{1.0}
  \caption{The MOS on naturalness in spontaneous-style speech of different models with 95\% confidence intervals. wo/label means the ground truth labels are not provided in inference phase.}
  \label{tab:mos}
  \centering
  \begin{tabular}{l|c|r} 
    \toprule
    \textbf{Model} &\textbf{MOS(w/label)}  &\textbf{MOS(wo/label)}\\
    \midrule
    Ground Truth & $4.483\pm0.061$ ~~~ & $4.516\pm0.059$ ~~~             \\
    FastSpeech 2 & $2.845\pm0.085$ ~~~ & $2.905\pm0.089$ ~~~ \\
    UCS* & $3.361\pm0.074$ ~~~ & $3.413\pm0.068$ ~~~\\
    Proposed & $\mathbf{3.561\pm0.073}$ ~~~ & $\mathbf{3.701\pm0.068}$ ~~~\\
    \bottomrule
  \end{tabular}
\end{table}

\begin{figure}[!htb]
	\centering
	\includegraphics[width=1.0\linewidth, height=0.3\linewidth]{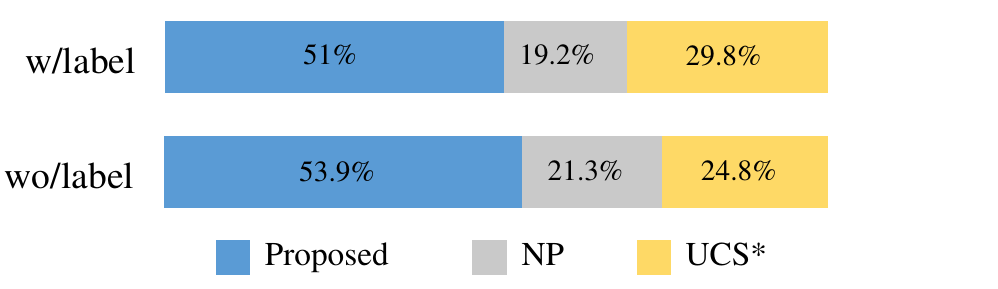}
	\caption{Results of the ABX test on naturalness in spontaneous-style speech between different models. NP means no preference.}
	\label{fig:abx1}
\end{figure}

In addition, ABX preference test is administered to ask subjects for their preferences regarding the naturalness and expressiveness of a pair of utterances generated by different models. Similarly, the tests were divided into two groups (with label and without label). We compare the proposed model with the UCS* model.
As shown in Fig.\ref{fig:abx1}, the preference rate of our proposed model exceeds UCS* by $21\%$ with label, and by $29\%$ without label.

MOS and ABX preference experiments show that our proposed method significantly outperforms the two baselines in synthesizing natural spontaneous-style speech, whether or not the given label is present in the inference phase.
Our proposed model outperforms the UCS* model when explicit labels are given, indicating that considering linguistic information can enhance the synthesis of spontaneous behavior.
In addition, we find that the proposed model outperforms *UCS significantly when no label are provided, indicating that semi-supervised pre-training can improve the model's ability to predict reasonable labels.

\subsection{Ablation Study}
Two ablation studies are conducted to demonstrate the efficacy of several techniques utilized in our proposed model, including the use of semi-supervised pre-training method and linguistics information.
The Comparative Mean Opinion Score (CMOS) was used to compare synthetic discourse in terms of predicting reasonable spontaneous behavior (wo/label) and the naturalness of spontaneous behavior (w/label).
The results are shown in Table \ref{tab:cmos}.
Eliminating the use of semi-supervised pre-training method (i.e., only training on the high-quality labeled corpus) results in $-0.343$ CMOS when the inference stage does not give labels. This demonstrates that the semi-supervised pre-training method significantly improves the model's ability to predict reasonable labels.
Moreover, we find that removing the linguistics-aware encoder results in $-0.248$ CMOS when the label is explicitly provided.
This indicates that the addition of the linguistic-aware encoder increases the ability of the models to simulate spontaneous behavior in speech.
Some audio samples are provided for listening\footnote{Sample: \href{https://thuhcsi.github.io/interspeech2023-spontaneousTTS/} 
{https://thuhcsi.github.io/interspeech2023-spontaneousTTS/}\label{demo}}.

\begin{table}[th]\footnotesize
\renewcommand{\arraystretch}{1.0}
  \caption{CMOS comparison for ablation study.}
  \label{tab:cmos}
  \centering
  \begin{tabular}{l|c} 
    \toprule
    \textbf{Model} &\textbf{CMOS} \\
    \midrule
    Proposed & $0$ ~~~ \\
    \quad -semi-supervised pre-training method (wo/label) & $-0.343$ ~~~ \\
    \quad -linguistics-aware encoder (w/label) & $-0.248$ ~~~ \\
    \bottomrule
  \end{tabular}
  \vspace{-1.5em}
\end{table}

\section{Conclusions}
\vspace{-0.6em}
In this paper, we propose a semi-supervised pre-training method conducted on a large-scale low-quality spontaneous dataset 
to increase the amount of spontaneous-style speech and spontaneous behavioral labels.
Text and speech information are considered for detecting spontaneous behaviors labels from speech in the process of semi-supervised learning.
In the proposed model, we utilize an linguistic-aware encoder to model the lingustic information in conversation.
Experimental results demonstrate that our proposed method achieves superior expressive speech synthesis performance by modeling spontaneous behavior in spontaneous-style speech and predicting reasonable spontaneous behavior from text.
.
\vspace{-0.8em}
\section{Acknowledgement}
This work is supported by National Natural Science Foundation of China (62076144), National Social Science Foundation of China (13\&ZD189), Shenzhen Science and Technology Program (WDZC20220816140515001) and Shenzhen Key Laboratory of next generation interactive media innovative technology (ZDSYS20210623092001004).

\newpage
\bibliographystyle{IEEEtran}
\bibliography{references}

\end{document}